\documentclass[preprint,
nofootinbib,
amsmath,amssymb,
 aps, physrev,
]{revtex4-2}

\usepackage{graphicx}\usepackage{dcolumn}\usepackage{bm}\usepackage{hyperref}\usepackage{physics} \usepackage{tensor} \usepackage[dvipsnames]{xcolor}
\usepackage{amsthm}
\usepackage{amssymb}
\usepackage{framed}

\newcommand{\Revise}[1]{{#1}}

\begin{document}

\title{Nonlinear tails in the Kerr black hole ringdown}

\author{Siyang Ling, Sam S. C. Wong}
\affiliation{Department of Physics, City University of Hong Kong,\\
  Tat Chee Avenue, Kowloon, Hong Kong SAR, China}
\email{siyaling@cityu.edu.hk}
\email{samwong@cityu.edu.hk}

\begin{abstract}
  Power law tails induced by nonlinearities of General Relativity (``sourced'' or ``nonlinear'' tails) were recently shown to dominate the late time waveform of Schwarzschild black hole ringdowns.
  We extend the analytical results regarding such nonlinear tails from Schwarzschild to Kerr black holes by studying the Teukolsky equation.
  Using a far field approximation to the radial Green's function, we analytically derived the tail power law for spin-weight $s \neq 0$, harmonic mode $(\ell m)$ and source decay $r^{-\beta}$.
  Specifically, we found a $t^{-\ell-\beta-s}$ tail for $\beta \leq 1$ or $\beta \geq \ell - s + 2$, and a $t^{-2\ell-3}$ Price tail for $2 \leq \beta \leq \ell - s + 1$.
  Our numerical tests were in exact agreement with these analytic results.
  We also demonstrate the dynamical formation of such nonlinear tails for a massless scalar by numerically solving the Teukolsky equation.
  In all numerical results, Kerr black hole nonlinear tails have the same power laws as that for Schwarzschild black holes, as expected from the Minkowski nature of the spacetime in the far field region. 
\end{abstract}

\maketitle

\newpage

\section{Introduction}
\label{sec:introduction}

Gravitational wave (GW) astronomy has witnessed rapid growth in popularity in recent years~\cite{Hughes:2001ch,Hughes:2002yy,Weinstein:2011kh,Cerdonio:2012ee,McWilliams:2019fng,Bailes:2021tot,Rosato:2025rtr}.
The current LIGO-VIRGO-KAGRA collaboration~\cite{KAGRA:2018plz,aLIGO:2020wna} and upcoming gravitational wave detectors, such as LISA~\cite{Baker:2019nia}, Taiji~\cite{Ruan:2018tsw}, TianQin~\cite{TianQin:2015yph}, DECIGO~\cite{Kawamura:2020pcg}, Cosmic Explorer~\cite{Reitze:2019iox} and Einstein Telescope~\cite{Punturo:2010zz} are detecting astrophysical gravitational waves at increasing levels of sensitivity~\cite{Borhanian:2022czq,Gupta:2023evt,Lu:2025nkt,Berti:2025hly,Perrone:2025zhy}.
These observations provide precision probes of black hole and neutron star mergers, offering unique insights into the merger processes, strong-field gravity, high energy physics and cosmology~\cite{Chamberlain:2017fjl,Martynov:2019gvu,Puecher:2022oiz,Callister:2024cdx,Puecher:2024kts,Perkins:2020tra,Gao:2022hsn,Bhagwat:2023jwv,Aly:2025dbx,Cardoso:2025npr}.
The prospect of detection has also motivated a variety of theoretical works on signatures of nonlinearities in ringdown waveforms~\cite{Nakano:2007cj,Ioka:2007ak,Sberna:2021eui,Redondo-Yuste:2023seq,Mitman:2022qdl,Cheung:2022rbm,Redondo-Yuste:2023ipg,Bourg:2024jme,Lagos:2024ekd,Perrone:2023jzq,Khera:2023oyf,Khera:2024bjs}.

One particular topic of interest is the nature of late time power law tails in the gravitational waveforms of binary black hole mergers.
Linear black hole perturbation theory (BHPT) predicts the existence of Price tails in the ringdown signals, with the tail power laws determined primarily by the spherical harmonic mode~\cite{Leaver:1986gd,Hod:1999ci,Barack:1998bv,Barack:1999st,Barack:1999ma,Hod:2000fh,Burko:2007ju,Gleiser:2007ti,Burko:2010zj,Konoplya:2011qq,Zenginoglu:2012us}.
However, recent numerical and analytical studies have demonstrated that late time gravitational waveforms are in fact dominated by ``nonlinear'' tails (or ``sourced'' tails), which arise from the nonlinearities inherent in General Relativity (GR)~\cite{Okuzumi:2008ej,Lagos:2022otp,Cardoso:2024jme,DeAmicis:2024eoy,Ma:2024hzq,Ling:2025wfv,Kehagias:2025xzm,Kehagias:2025tqi,Ianniccari:2025avm,Ianniccari:2025avm,Bishoyi:2026cpg}.
Furthermore, similar nonlinear mechanisms can lead to tails in scalar and electromagnetic fields, making nonlinear tails a topic of broader interest in astrophysics~\cite{Bizon:2008iz,Aly:2024hfz,Aly:2024kmx,He:2025ydh,Shao:2026hnd}.
The detection of these tails will provide a direct probe of strong-field gravity and could constrain potential deviations from GR.

While astrophysical merger remnants are expected to be Kerr (rotating) black holes~\cite{Owen:2009sb}, the majority of analytical work on nonlinear tails has focused on Schwarzschild (non-rotating) black holes.
Although some attempts have been made on Kerr black hole nonlinear tails~\cite{Perrone:2025zhy,Bishoyi:2026cpg}, full analytical understanding of Kerr nonlinear tails remains elusive.
In this work, we aim to bridge this gap by performing an analytical and numerical study of Kerr black hole nonlinear tails.
We employ the Teukolsky formalism~\cite{Teukolsky:1973ha}, a unified framework for evolving perturbations of any spin-weight (scalar, electromagnetic, or gravitational) in Kerr spacetime, so that a coherent picture across different fields can be obtained.

A key insight from recent studies on Schwarzschild black holes is that nonlinear tails are predominantly generated in the far field regime ($M / r \ll 1$, in units of $G=1$), where the spacetime is approximately Minkowski, by the propagation of outgoing wavepackets (e.g. quasinormal modes)~\cite{Okuzumi:2008ej,Lagos:2022otp,Cardoso:2024jme,Kehagias:2025xzm,Kehagias:2025tqi,Ling:2025wfv}.
In the far field regime of a Kerr black hole ($M / r \ll 1$ and $a / r \ll 1$), the spacetime also approximates Minkowski spacetime, similar to the case of Schwarzschild black holes.
We thus expect the mechanism for generating nonlinear tails to be analogous in both Kerr and Schwarzschild black holes.
Indeed, our numerical results show that spin-dependent background terms (those dependent on $a$) have no effect on tail power-law indices.

In this work, we apply the Green's function methods that has been successful in the Schwarzschild case to the Teukolsky equation.
Working in spin-weighted spherical harmonic basis (${}_s Y_{\ell m}$), we derive the approximate Green's function for the Teukolsky equation in the far field limit, and compute the resulting power law tails for an outgoing source with decay $r^{-\beta}$.
We find analytically that the power law is $t^{-\ell-\beta-s}$ for $\beta \leq 1$ or $\beta \geq \ell -s + 2$, where $s$ is the spin-weight.
We verify this result by numerical evaluations for a range of parameters of ($\ell$, $\beta$, $s$), and a simulation on the dynamical formation of power law tails sourced by a massless scalar around a Kerr black hole.

\section{Black hole perturbations on Kerr background}
\label{sec:theory}

We follow Ref.~\cite{Chandrasekhar:1985kt}'s convention for the Kerr metric and Teukolsky equation.
In Boyer-Lindquist coordinates, a black hole with mass $M$ and angular momentum $a M$ is given by the Kerr metric:
\begin{align}
  &\dd{s}^2 = -\left(1 - \frac{2Mr}{\Sigma}\right) \dd{t}^2 
  - \frac{4aMr \sin^2\theta}{\Sigma} \dd{t} \dd{\phi} 
  + \frac{\Sigma}{\Delta} \dd{r}^2 \nonumber \\
  &\quad + \Sigma \dd{\theta}^2 
  + \left(r^2 + a^2 + \frac{2a^2 M r \sin^2\theta}{\Sigma}\right) \sin^2\theta \dd{\phi}^2 ,
  \label{eq:kerr}
\end{align}
where $\Sigma \equiv r^2 + a^2 \cos^2\theta$, $\Delta \equiv r^2 - 2Mr + a^2$ and $r_\pm \equiv M \pm \sqrt{M^2 - a^2}$.
The black hole event horizon is at $r_+$.
Also, the tortoise coordinate $r_*$ is defined as
\begin{align}
  r_* &= r + \frac{r_+^2 + a^2}{r_+ - r_-} \ln(r - r_+)
        + \frac{r_-^2 + a^2}{r_+ - r_-} \ln(r - r_-) \;.
\end{align}
This coordinate satisfies $\dv*{r}{r_*} = \Delta / (r^2 + a^2)$ and maps $r \in (r_+, \infty)$ to $r_* \in (-\infty,\infty)$.

Instead of the spheroidal harmonics, we choose the spin-weighted spherical harmonics ${}_s Y_{\ell m}$ as a basis for the angular functions.
When $s=0$, we simply write $Y_{\ell m} = {}_0 Y_{\ell m}$.
The spin-weighted spherical harmonics form a complete orthonormal basis~\cite{Goldberg:1966uu,Thorne:1980ru}
\begin{align}
  & \int_0^{2\pi}\dd{\phi} \int_{-1}^{1} \dd{\cos(\theta)} {}_sY_{\ell m}^\ast(\theta,\phi) {}_sY_{\ell'm'}(\theta,\phi) = \delta_{\ell \ell'} \delta_{mm'} \nonumber \\
  & \sum_{(\ell m)} {}_sY_{\ell m}^\ast(\theta,\phi) {}_sY_{\ell m}(\theta',\phi') = \delta(\cos(\theta) - \cos(\theta')) \delta(\phi-\phi') \;.
\end{align}

\subsection{Teukolsky equation}
\label{sec:teukolsky_equation}

The evolution of field perturbations around Kerr black holes is given by the Teukolsky equation.
The Teukolsky equation for generic field perturbation $\psi(t,r,\theta,\phi)$ and generic source $\mathcal{S}$ is written in terms of the Teukolsky operator ${}_s\mathcal{T}$ as~\cite{Teukolsky:1972my,Teukolsky:1973ha,Teukolsky:1974yv}
\begin{align}
  \label{eq:teukolsky}
  {}_s\mathcal{T} \psi &= \mathcal{S} \qq{where} \nonumber \\
  {}_s\mathcal{T} &\equiv \left[\frac{(r^2+a^2)^2}{\Delta}
                  - a^2 \sin^2\theta\right] \pdv[2]{t} 
                  + \frac{4 M a r}{\Delta} \pdv{t} \pdv{\phi} \nonumber \\
                     &\quad + \left[\frac{a^2}{\Delta} - \frac{1}{\sin^2\theta}\right] \pdv[2]{\phi} 
                       - \Delta^{-s} \pdv{r} (\Delta^{s+1} \pdv{r})
                       - \frac{1}{\sin\theta} \pdv{\theta} (\sin\theta \pdv{\theta}) \nonumber \\
                     &\quad - 2 s \left[ \frac{a(r - M)}{\Delta} + \frac{i \cos\theta}{\sin^2\theta} \right] \pdv{\phi}
                       - 2 s \left[ \frac{M(r^2 - a^2)}{\Delta} - r - i a \cos\theta \right] \pdv{t} \nonumber \\
                     &\quad + (s^2 \cot^2\theta - s) \;.
\end{align}
Here, the spin-weight $s$ and the source $\mathcal{S}$ depends on the type of perturbation being treated.
The Teukolsky equation describes the dynamics of scalar ($s=0$, reducing to the Regge-Wheeler equation for scalar), vector/electromagnetic ($s = \pm 1$), and tensor/Weyl scalar ($s = \pm 2$) perturbations~\cite{Teukolsky:1973ha}.

To facilitate subsequent analyses,  we introduce a change of variable 
\begin{align}
    \tilde{\psi} \equiv (\Delta^s r) \psi.
\end{align}
This change of variable is motivated by the following: it is well-known that, upon separation in terms of spin-weighted spheroidal harmonics, the radial solutions to the Teukolsky equation have asymptotic form $R_{\ell m}^{\mathrm{in}}(r,\omega) \propto \Delta^{-s} e^{-i \omega r_*}$ as $r \to r_+$ and $R_{\ell m}^{\mathrm{out}}(r,\omega) \propto r^{-1-2s} e^{i \omega r_*}$ as $r_* \to \infty$~\cite{Casals:2016soq}.
The power law dependence of $R_{\ell m}^{\mathrm{in},\mathrm{out}}$ on $r$ leads to either very large or small numerical values of $\psi$ at $r_* \to \pm \infty$.
To reduce the range of numerical amplitudes of the variables, we instead use $\tilde{\psi} \equiv (\Delta^s r) \psi$, which has asymptotic behavior $e^{\pm i \omega r_*}$ at $r_* \to \pm \infty$.
We also define the corresponding Teukolsky operator ${}_s\tilde{\mathcal{T}}$ and source $\tilde{\mathcal{S}}$:
\begin{align}
  {}_s\tilde{\mathcal{T}} &\equiv \left[\frac{(r^2+a^2)^2}{\Delta} - a^2 \sin^2\theta\right]^{-1} (\Delta^s r) {}_s\mathcal{T} (\Delta^s r)^{-1} \nonumber \\
\tilde{\mathcal{S}} &\equiv \left[\frac{(r^2+a^2)^2}{\Delta} - a^2 \sin^2\theta\right]^{-1}
                        (\Delta^s r) \mathcal{S} \;,
\end{align}
and the Teukolsky equation becomes ${}_s\tilde{\mathcal{T}}\tilde{\psi} = \tilde{\mathcal{S}}$ in these new notations.
Note that the coefficient for $\partial_t^2$ is normalized to $1$ in ${}_s\tilde{\mathcal{T}}$.

To isolate the radial evolution, we expand $\tilde{\psi}(t,r,\theta,\phi)$ in terms of the spin-weighted spherical harmonics ${}_sY_{\ell m}(\theta,\phi)$~\cite{Goldberg:1966uu,Thorne:1980ru}, and project the Teukolksy equation \eqref{eq:teukolsky} onto this basis \footnote{Spin-weighted spherical harmonics are chosen over spheroidal harmonics for numerical tractability; the primary complication introduced by this choice is $\ell$-mode mixing when $a \neq 0$.}:
\begin{align}
  \label{eq:psi_lm_pde}
  \tilde{\psi}(t,r,\theta,\phi) &= \sum_{\ell=|s|}^{\infty}\sum_{m=-\ell}^{\ell} \tilde{\psi}_{\ell m}(t,r)  {}_sY_{\ell m}(\theta,\phi) \nonumber\\
  ({}_s\tilde{\mathcal{T}} \tilde{\psi})_{\ell m}(t,r) &= \tilde{\mathcal{S}}_{\ell m}(t,r) \;,
\end{align}
where
\begin{align}
  ({}_s\tilde{\mathcal{T}} \tilde{\psi})_{\ell m} &= \partial_t^2 \tilde{\psi}_{\ell m} + \sum_{(\ell'm')} C_{\ell'm'\ell m} \tilde{\psi}_{\ell'm'}
     + C_{\ell'm'\ell m}^{(t)} \partial_t \tilde{\psi}_{\ell'm'} \nonumber \\
   &  + C_{\ell'm'\ell m}^{(r)} \partial_r \tilde{\psi}_{\ell'm'}
     +  C_{\ell'm'\ell m}^{(rr)} \partial_r^2 \tilde{\psi}_{\ell'm'} \;.
\end{align}
Here, the coefficients $C_{\ell m\ell'm'}^{(\ast)}(r)$ are implicitly defined by applying ${}_s\tilde{\mathcal{T}}$ on the ${}_s Y_{\ell m}$'s, and are fixed by the Kerr parameters $M$ and $a$.
Note that $\partial_t^2$ has unit coefficient by normalization of ${}_s\tilde{\mathcal{T}}$ and orthonormality of the ${}_s Y_{\ell m}$'s.
Moreover, axis-symmetry of the Teukolsky equation implies $C_{\ell m\ell'm'}^{(\ast)} = 0$ for $m \neq m'$; if $a=0$, then spherical symmetry further implies $C_{\ell m\ell'm'}^{(\ast)} = 0$ for any $(\ell m) \neq (\ell' m')$, and ${}_s \tilde{\mathcal{T}}$ is diagonalized by the ${}_s Y_{\ell m}$'s.
These equations form a complete system of 1+1D partial differential equations (PDEs) for $\tilde{\psi}_{\ell m}(t,r)$.

\subsection{Green's function in Schwarzschild and far field limit}
\label{sec:far_field_limit}

It was argued in Ref.~\cite{Ling:2025wfv} that, for Schwarzschild black holes, nonlinear tails due to outgoing wavepackets are primarily sourced far from the horizon ($r \gg M$), where the spacetime is approximately Minkowski.
Intuitively speaking, since $M / r \ll 1$ implies $\abs{a} / r \ll 1$, Kerr spacetime \eqref{eq:kerr} is also approximately Minkowski far away from the horizon, and the spacetime background on which Kerr nonlinear tails are sourced is similar to that for Schwarzschild nonlinear tails.
Consequently, Kerr nonlinear tails are expected to have the same power laws and amplitudes as that for Schwarzschild nonlinear tails.
In this section, we discuss some properties of the Green's function for the Teukolsky equation \eqref{eq:psi_lm_pde} in the Schwarzschild ($a = 0$) and the far field limit ($a,M = 0$).
These results generalize the Green's function analyses in Ref.~\cite{Cardoso:2024jme,Ling:2025wfv,Kehagias:2025xzm} to nonzero spin-weights.

The Teukolsky operator \eqref{eq:psi_lm_pde} for $a = 0$ is given by:
\begin{align}
  \label{eq:Sch_T_tilde_slm}
  {}_s\tilde{\mathcal{T}}_{\ell m} &= \pdv[2]{t} + \frac{2 s (r - 3 M)}{r^2} \pdv{t} + \frac{r - 2 M}{r^3} (\ell-s)(\ell+s+1) \nonumber \\
                                   &\quad + \frac{r - 2 M}{r^4} \left( (2 M - r) r^2 \pdv[2]{r} - 2 r (M + s (M - r)) \pdv{r}  + 2 M (1 + s) \right) \;,
\end{align}
where angular part in $ {}_s\tilde{\mathcal{T}}$ is handled by the $\bar{\eth} \eth$ operator, with identity $\bar{\eth} \eth {}_s Y_{\ell m} = -(\ell-s)(\ell+s+1) {}_s Y_{\ell m}$~\cite{Goldberg:1966uu}.\footnote{The $\bar{\eth} \eth$ operator is given by $\bar{\eth} \eth = \partial_\theta^2 + \cot\theta \partial_\theta + \csc^2\theta \partial_\phi^2 + 2 i s \cot\theta \csc\theta \partial_\phi - (s^2 \cot^2\theta -s)$.}
Note that ${}_s\tilde{\mathcal{T}}_{\ell m}$ depends on $\ell$ but not on $m$, as guaranteed by spherical symmetry.
The retarded Green's function $G_{s \ell m}(r,t;r',t')$ for ${}_s \tilde{\mathcal{T}}_{\ell m}$ is defined by
\begin{align}
  {}_s\tilde{\mathcal{T}}_{\ell m} G_{s \ell m}(r,t;r',t') = - \delta(r-r') \delta(t-t')
\end{align}
for each fixed harmonic mode $(\ell m)$.

The retarded and advanced Green's function $G_{s \ell m}^{\mathrm{ret/adv}}$ admits an useful exact identity under $s$ inversion $s \to -s$:
\begin{align}
  \label{eq:G_s_inversion_symmetry}
  G_{s \ell m}^{\mathrm{adv}}(r,t;r',t') = \left( \frac{\Delta(r)}{\Delta(r')} \right)^s G_{(-s) \ell m}^{\mathrm{ret}}(r,-t;r',-t') \;,
\end{align}
or
\begin{align}
  G_{s \ell m}^{\mathrm{ret}}(r,t;r',t') = \left( \frac{\Delta(r)}{\Delta(r')} \right)^s G_{(-s) \ell m}^{\mathrm{adv}}(r,-t;r',-t') \;,
\end{align}
where $\Delta(r) = r^2 - 2 M r$ .
This exact identity holds even near the horizon.
To show this, let $T$ be the time reversal operator, $[Tf](t,r) = f(-t,r)$, and $[\Delta^s f](r) = \Delta(r)^s f(r)$.
One can check the following operator identity
\begin{align}
  T \Delta^{-s} {}_s\tilde{\mathcal{T}} \Delta^s T = {}_{-s}\tilde{\mathcal{T}} \;.
\end{align}
Note that\footnote{This identity can be in fact be generalized to $a \neq 0$, as a symmetry in the radial Teukolsky functions $R_{s \ell m \omega}$.}
\begin{align}
  & [T \Delta^{-s} {}_s\tilde{\mathcal{T}} \Delta^s T G_{(-s) \ell m}^{\mathrm{ret}}](r,t;r',t')
  = [{}_{-s}\tilde{\mathcal{T}}_{\ell m} G_{(-s) \ell m}^{\mathrm{ret}}](r,t;r',t')
  = - \delta(r-r') \delta(t-t') \nonumber \\
  & [{}_s\tilde{\mathcal{T}} \Delta^s T G_{(-s) \ell m}^{\mathrm{ret}}](r,t;r',t')
    = - \Delta(r')^s \delta(r-r') \delta(-t-t') \;.
\end{align}
By definition, $[\Delta^s T G_{(-s) \ell m}^{\mathrm{ret}}](r,t;r',t')$ is equal to $\Delta(r')^s G_{s \ell m}^{\mathrm{adv}}(r,t;r',-t')$ plus some homogeneous solution to ${}_s\tilde{\mathcal{T}} f = 0$.
Since $G_{(-s) \ell m}^{\mathrm{ret}}$ is defined as the retarded Green's function, it is necessary that $[\Delta^s T G_{(-s) \ell m}^{\mathrm{ret}}](r,t;r',t') = \Delta(r)^s G_{(-s) \ell m}^{\mathrm{ret}}(r,-t;r',t') = 0 =  \Delta(r')^s G_{s \ell m}^{\mathrm{adv}}(r,t;r',-t') $ for $-t-t'<0$.
These two conditions imply that $\Delta^s T G_{(-s) \ell m}^{\mathrm{ret}}$ is proportional to the advanced Green's function, and hence Eq.~\eqref{eq:G_s_inversion_symmetry}.

In the far field approximation ($a/r,M/r \ll 1 $), the radial Teukolsky equation in frequency space is approximated as 
\begin{align}
  \label{eq:far_field_equation}
 0 &= \pdv[2]{{}_s\tilde{\psi}_{\ell m}}{r} - \frac{2 s}{r} \pdv{{}_s\tilde{\psi}_{\ell m}}{r} 
  - \frac{1}{r^2} (\ell-s)(\ell+s+1) {}_s\tilde{\psi}_{\ell m}
  - (- \omega^2 - i \omega \frac{2s}{r}) {}_s\tilde{\psi}_{\ell m} \;,
\end{align}
which admits exact solutions in terms of Whittaker functions ($  \tilde{\psi}_{\infty_\pm}(\omega, r) \approx r^s W_{\mp s,\ell+1/2}(\mp 2 i \omega r)$).
We used these solutions to construct approximate Green's function via similar methods as in Ref.~\cite{Ling:2025wfv}.
For example, for $\ell = 2$, the far field approximation yields 
\begin{align}
  \label{eq:far_field_greens_function}
  &\bar{G}_{s (\ell = 2) m}(r,t;r',t') \nonumber\\ \approx 
      &\begin{cases}
        \frac{(r + r' + (t-t'))^4}{32 r^4} & s = -2 \\
        \frac{(r + r' + (t-t'))^2(r^2 - r r' + (r')^2 - (t-t')^2)}{8 r^3 r'}  & s = -1 \\
        \frac{3 r^4 + 2 r^2 ((r')^2 - 3 (t-t')^2) + 3( (r')^2 - (t-t')^2)^2}{16 r^2 (r')^2} & s = 0 \\
        \frac{(r + r' - (t-t'))^2(r^2 - r r' + (r')^2 - (t-t')^2)}{8 r (r')^3} & s = 1 \\
        \frac{(r + r' - (t-t'))^4}{32 (r')^4} & s = 2
      \end{cases} \nonumber \\
    & {\rm for} \abs{r_* - r_*'} < t-t' < r_* + r_*'\;.
\end{align}
where $\bar{G}_{s \ell m}$ is the Green's function without the causal theta function, i.e. $G_{s \ell m} = \bar{G}_{s \ell m} \Theta(t-t'-|r_*-r_*'|)$.
For $s = 0$, the Green's functions are identical to those derived for the Regge-Wheeler equation in Ref.~\cite{Barack:1999ma,Ling:2025wfv}, consistent with the fact that the Teukolsky equation reduces to the Regge-Wheeler equation in this case.
We have also computed Green's functions via direct numerical integration, and found that the approximations for $\ell=0,1,2$ are accurate within their regime of validity.
Also note that these approximations explicitly satisfy identity \eqref{eq:G_s_inversion_symmetry}.

Finally, we discuss the far field approximation in further detail.
At large $r$, the Kerr metric \eqref{eq:kerr} and the corresponding Teukolsky operator are asymptotically:
\begin{align}
  \dd{s}^2 &= \left[- \dd{t}^2 + \dd{r}^2 + r^2 \dd{\theta}^2  + r^2 \sin^2\theta \dd{\phi}^2 \right] \times \left( 1 + \order{1/r} \right) \nonumber \\
  ({}_s\tilde{\mathcal{T}} \tilde{\psi})_{\ell m} &= \partial_t^2 \tilde{\psi}_{\ell m} - \partial_r^2 \tilde{\psi}_{\ell m } + \frac{2s}{r} \partial_t \tilde{\psi}_{\ell m } + \frac{2s}{r} \partial_r \tilde{\psi}_{\ell m } + \frac{(\ell - s)(\ell + s + 1)}{r^2} \tilde{\psi}_{\ell m} \nonumber \\
           &\qquad + \order{a \tilde{\psi} /r^3, M \tilde{\psi} /r^3, M \partial_t\tilde{\psi} /r^2, M \partial_r\tilde{\psi} /r^2} \;. 
\end{align}
The written terms above are exactly those given by a spin-$s$ wave equation in Minkowski spacetime, which contain no mode-mixing, and there is no subleading terms in the corresponding Green's function.
Since $a$ and $M$ are absent at leading order, when solving the equation $\tilde{\mathcal{T}}\tilde{\psi} = \tilde{\mathcal{S}}$ perturbatively for a fixed nonzero source $\tilde{\mathcal{S}}$, corrections to the solution are of order $\order{M/r,a/r}$.

\subsection{Response to outgoing source}
\label{sec:teukolsky_response}
In this section, we compute, in the far field approximation, the response to a source $\tilde{\mathcal{S}}$ of the form:
\begin{align}
  \label{eq:outgoing_source}
  {}_s\tilde{\mathcal{T}} \tilde{\psi} = \tilde{\mathcal{S}} \qq{where}
  \tilde{\mathcal{S}} = \frac{F((t-t_i) - (r - r_i))}{r^\beta} {}_s Y_{\ell m}
\end{align}
for $t \geq t_i$, $r \geq r_i$, and $F$ is supported on $[-\sigma, \sigma]$ only.
Physically, this source represents an outgoing wavepacket moving at the speed of light, with its waveform fixed by $F$ and its amplitude decaying like $r^{-\beta}$.
Sources of this form are motivated by nonlinear black hole dynamics: outgoing quasinormal modes are generally formed during black hole ringdowns, and quadratics of these outgoing modes act as effective sources for higher-order perturbations in the framework of black hole perturbation theory~\cite{Cardoso:2024jme,Ling:2025wfv}.
Consequently, the power laws derived from linear response to appropriate sources are, to leading order, identical to those coming from nonlinearities of GR.

\begin{figure}[t]
  \centering
  \includegraphics[width=0.48\textwidth]{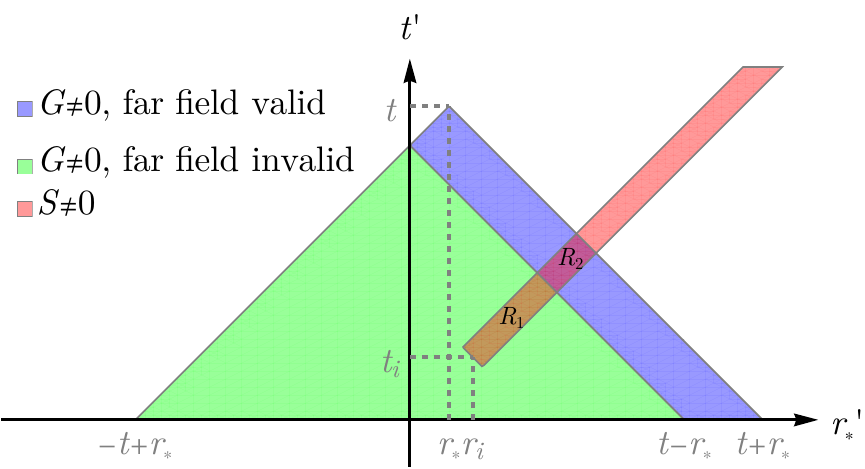} 
  \caption{Integration region for the response integral \eqref{eq:response_integral}.
    The Green's function is nonzero in both green and blue region, but the far field approximation is only valid in the blue region.
    Rectangular regions $R_1$ and $R_2$ have nonzero contributions to the integral, and analytical evaluation is possible only in $R_2$.
    For $s>0$, contribution from $R_2$ dominates the full integral.}
  \label{fig:integration_region}
\end{figure}

In order to find the response, we want to evaluate the integral
\begin{align}
  \label{eq:response_integral}
  \tilde{\psi}_{\ell m}(t,r_*) &= \int_{-\infty}^{t} \int_{r_* - (t-t')}^{r_* + (t-t')} G_{s \ell m}(r_*,t;r_*',t') \tilde{\mathcal{S}}_{\ell m}(r_*',t') \dd{r_*'} \dd{t'} \;.
\end{align}
See Fig.~\ref{fig:integration_region} for an illustration of the integration region.
We perform this integral in frequency space:
\begin{align}
  \label{eq:frequency_space_response_integral}
  \tilde{\psi}_{\ell m}(t,r_*) &= \frac{1}{2\pi i} \int_{ic-\infty}^{ic+\infty} e^{-i\omega t} \tilde{\psi}_{\ell m}(\omega,r_*) (-i)\dd{\omega} \nonumber \\
  \tilde{\psi}_{\ell m}(\omega,r_*)  &= \int_{r_i}^{\infty} g_{s \ell m}(r_*,r_*',\omega) \tilde{\mathcal{S}}_{\ell m}(r_*',\omega) \dd{r_*'} \;,
\end{align}
where $g_{s \ell m}$ is the frequency space Green's function, and variables with $\omega$ arguments (e.g. $\tilde{\mathcal{S}}_{\ell m}(\omega, r_*)$) are Fourier transforms of the corresponding quantities:
\begin{align}
  G_{s \ell m}(r_*, t; r_*', t')  &= \frac{1}{2 \pi i} \int_{ic - \infty}^{ic + \infty} e^{- i \omega (t-t')} g_{s \ell m}(r_*,r_*',\omega) (- i) \dd{\omega} \nonumber \\
  \tilde{\mathcal{S}}_{\ell m}(\omega,r') &= (r')^{-\beta} e^{i\omega (r_*' + t_i - r_i)} F(\omega) \;.
\end{align}
Note that the causality condition is implicitly satisfied by choosing the integral contour such that $c$ is above all poles of $g_{s \ell m}$.

To evaluate this integral, we construct $g_{s \ell m}$ using the following recipe:
\begin{align}
  g_{s \ell m}(r,r',\omega) &\equiv \frac{\tilde{\psi}_{\infty_+}(\omega, r_>) \tilde{\psi}_{0}(\omega, r_<) }{W[\tilde{\psi}_{\infty_+}, \tilde{\psi}_{0}](r')} \nonumber \\
  \tilde{\psi}_{\infty_+}(\omega, r) &\equiv r^s W_{-s,\ell+1/2}(-2 i \omega r) \nonumber \\
  \tilde{\psi}_{\infty-}(\omega, r) &\equiv r^s W_{s, \ell+1/2}(2 i \omega r) \nonumber \\
  \tilde{\psi}_0(\omega, r) &\equiv \Gamma(\ell + 1 + s) \tilde{\psi}_{\infty_+}(\omega, r) +
                              (-1)^{\ell + 1} \Gamma(\ell + 1 - s) \tilde{\psi}_{\infty_-}(\omega, r) \nonumber \\
  W[\tilde{\psi}_{\infty_+}, \tilde{\psi}_0](r) &= \Gamma(\ell + 1 - s) (-1)^{\ell + s} 2 i \omega r^{2s} \;.
\end{align}
Here, $W_{k,m}(z)$ is the Whittaker function, and $\tilde{\psi}_{\infty_\pm}$ are the outgoing and ingoing solutions of the far field equation \eqref{eq:far_field_equation} at $r \to \infty$.
One can explicitly check that $g_{s \ell m}$ and $\tilde{\psi}_0(\omega, r_*)$ are regular around $\omega = 0$, with $\tilde{\psi}_0(\omega, r_*) = \order{\omega^{\ell+1}}$.
In fact, $\tilde{\psi}_0$ is chosen to be the linear combination of $\tilde{\psi}_{\infty\pm}$ such that its matched solution is ingoing at the horizon.\footnote{$\tilde{\psi}_0$ must be the regular ingoing solution at the horizon, as required by causality. Consider the frequency space Teukolsky equation ${}_s \tilde{\mathcal{T}}_{\ell m} \tilde{\psi} = 0$; in the $\omega = 0$ limit, its general solution at $r \to \infty$ splits into two branches, $\sim r^{s-\ell}$ and $\sim r^{\ell + s + 1}$. Furthermore, the authors of Ref.~\cite{Hui:2021vcv} showed that any static black hole solution that contains nonzero decaying branch ($\sim r^{s-\ell}$) must be irregular at the horizon. Our stated $\tilde{\psi}_0$ as a linear combination of $\tilde{\psi}_{\infty\pm}$ eliminates the $r^{s-\ell}$ branch, so it is matched to the horizon-ingoing exact solution in the far field regime.}
To proceed, we assume $r' > r$ for concreteness, then $g_{s \ell m}$ is asymptotically
\begin{align}
  g_{s \ell m}(r_*, r_*', \omega) = \frac{ \tilde{\psi}_{0}(\omega, r) }{\Gamma(\ell + 1 - s) (-1)^{\ell + s} 2 i \omega } (r')^{-s} W_{-s, \ell+1/2}(-2i\omega r') \;,
\end{align}
and the $r_*'$ integral is explicitly given by:
\begin{align}
  \tilde{\psi}_{\ell m}(\omega,r_*) = \frac{ \tilde{\psi}_{0}(\omega, r) e^{i\omega(t_i - r_i)} F(\omega) }{\Gamma(\ell + 1 - s) (-1)^{\ell + s} 2 i \omega }  \int_{r_i}^\infty (r')^{-(\beta + s)} e^{i\omega r_*'} W_{-s, \ell+1/2}(-2i\omega r')  \dd{r_*'} \;.
\end{align}

We first make the substitution $x = -2i\omega r'$ for the above Mellin integral:
\begin{align}
  I_{s,\beta}(x_i) \equiv (-2i\omega)^{\beta + s - 1} \int_{x_i}^\infty x^{-(\beta + s)} e^{-x/2} W_{-s, \ell+1/2}(x)  \dd{x} \;.
\end{align}
Our regime of interest is $\abs{x_i} \ll 1$, corresponding to $t \gg r_i$.
Depending on the choice of $(s,\ell,\beta)$, $I_{s,\beta}$ can be sensitive to the integration boundary $x_i \equiv -2 i \omega r_i$, and be divergent for $x_i = 0$.
In particular, if the integrand contains a $x^{-1}$ piece in its Laurent expansion, then the integral would have a logarithmic dependence on $x_i$:
\begin{align}
  I_{s,\beta}(x_i) \approx
  \begin{cases}
    (-2i\omega)^{\beta + s - 1} \ln(-2i\omega r_i) \times \mathrm{constant} + \hdots & \textrm{if integrand has $x^{-1}$ term} \\
    \sum_{j} c_j \omega^j & \textrm{if integrand has no $x^{-1}$ term}
  \end{cases}
  \;.
\end{align}
We will see that this logarithmic piece ultimately determines the contribution to the $\omega$ integral.
One can explicitly check that integrand $x^{-(\beta + s)} e^{-x/2} W_{-s, \ell+1/2}(x)$ does not have a $x^{-1}$ piece exactly when $2 \leq \beta \leq \ell - s + 1$, and it is for these values that the logarithmic piece vanishes.\footnote{Another way to understand this is to evaluate $I_{s,\beta}(0)$ and consider its analytic continuation in $\beta$:
  \begin{align}
    I_{s,\beta}(0) =
    (-2i\omega)^{\beta + s - 1} \frac{\Gamma(1-\ell-s-\beta)\Gamma(2+\ell-s-\beta)}{\Gamma(2-\beta)} \;.
  \end{align}
  The above function have poles for certain choices of $(s,\ell,\beta)$.
  One could show that the $\omega$ integral should be proportional to the residue of $I_{s,\beta}$ at $\beta$.
  For $2 \leq \beta \leq \ell - s + 1$, $I_{s,\beta}$ is regular at $\beta$, so the residue (and hence the nonlinear tail) vanishes.
}

For the part outside of the integral, we find that around $\omega = 0$:
\begin{align}
  \frac{ \tilde{\psi}_{0}(\omega, r) e^{i\omega(t_i - r_i)} F(\omega) }{\Gamma(\ell + 1 - s) (-1)^{\ell + s} 2 i \omega } =
    e^{i\omega(t_i - r_i)} F(0) \frac{\Gamma(\ell+1+s)}{\Gamma(2\ell+2)} r^{\ell+s+1} (-2i\omega)^\ell + \order{\omega^{\ell+1}} \;,
\end{align}
where we have expanded $F(\omega) = F(0) + \order{\omega}$ by assuming that $F$ has Gaussian profile.
We can now combine above results and evaluate $\tilde{\psi}_{\ell m}(t,r_*)$ from $\tilde{\psi}_{\ell m}(\omega,r_*)$.
For $\beta \leq 1$ or $\beta \geq \ell-s+2$, $\tilde{\psi}_{\ell m}$ contains a logarithmic piece of the form:
\begin{align}
  \tilde{\psi}_{\ell m}(\omega, r_*) \propto \omega^{\ell + \beta + s - 1} \ln(\omega) \;,
\end{align}
and the $\omega$ integral can then be evaluated to yield $t^{-\ell-\beta-s}$ if $\ell + \beta + s \neq 0$.
For $2 \leq \beta \leq \ell - s + 1$, there is no logarithmic piece and the branch contribution vanishes, so there is no nonlinear tail for these values.
For example, for $s=-1,\ell=1$, the tail vanishes for $\beta = 2, 3$.
Finally, we note the special case $\ell + \beta + s = 0$, wherein $\tilde{\psi}_{\ell m}(\omega, r_*)$ simplifies to a simple pole, and we get $\tilde{\psi}_{\ell m} \propto t^0$, consistent with the general rule $t^{-\ell-\beta-s}$.

In summary, for fixed {\bf integer} $\beta$ and $\ell$, the tail power laws are given by:
\begin{align}
  \label{eq:power_law_prediction}
  & t^{-\beta-\ell} \qq{for} s = 0, \quad \beta \leq 1 \nonumber \\
  & t^{-2\ell-2} \qq{for} s = 0, \quad 2 \leq \beta \leq \ell + 2 \nonumber \\
  & t^{-\beta-\ell-s} \qq{for} s \neq 0, \quad \beta \leq 1 \qq{or} \beta \geq \ell - s + 2 \nonumber \\
  & t^{-2\ell-3} \qq{for} s \neq 0, \quad 2 \leq \beta \leq \ell - s + 1 \;.
\end{align}
Here, the $s=0$ case is identical to the ones given in \cite{Ling:2025wfv}, since equation \eqref{eq:Sch_T_tilde_slm} for $s=0$ is identical to the Regge-Wheeler equation.
For the last row, the Price tail dominates, so the overall power law is $t^{-2\ell-3}$.

\section{Numerical results}
\label{sec:numerics}
In this section, we present numerical results for the sourced solution of the Teukolsky equation. Unlike the analytic treatment in Sec.~\ref{sec:theory}, the numerical computation does not rely on any far field or other simplifying approximations, and therefore captures the full dynamics within the adopted setup. This provides a direct test of the analytic understanding developed above. As we will show, in the far-from-black-hole region, the numerical results agree very well with the analytic predictions, thereby confirming the validity of the far field approximation and the physical picture inferred from it.

Our code is released at \url{https://github.com/hypermania/BlackholePerturbations}.

\subsection{Outgoing source}
\label{sec:outgoing_source}
We numerically solved the Teukolsky equation \eqref{eq:psi_lm_pde} with outgoing source profile \eqref{eq:outgoing_source}.
The source was chosen with parameters similar to that in Ref.~\cite{Ling:2025wfv}:
\begin{align}
 t_i/(2M) = 10, r_i/(2M) = 10, F(x) = \frac{1}{\sqrt{2\pi}\sigma} \exp(-\frac{x^2}{2\sigma^2}) \qq{where} \sigma/(2M) = 0.5 \;.
\end{align}
The source harmonic mode was chosen such that $m_{\mathrm{source}} = \ell_{\mathrm{source}}$, and the source spin-weight $s$ matches that of the Teukolsky equation.
We took zero initial condition for $\tilde{\psi}$ at $t=0$, and evolved $\tilde{\psi}$ up to $t=1000$.
We numerically solved the equations for combinations $-2 \leq s \leq 0$, $\abs{s} \leq \ell_{\mathrm{source}} \leq 2$, $0 \leq \beta \leq 2$, and $a/M = 0, 0.1, 0.2, 0.4, 0.8$.
All harmonic modes with $\ell \leq \ell_{\mathrm{max}} = 5$ are included in the evolution.

\begin{figure}[t]
  \centering \includegraphics[width=0.48\textwidth]{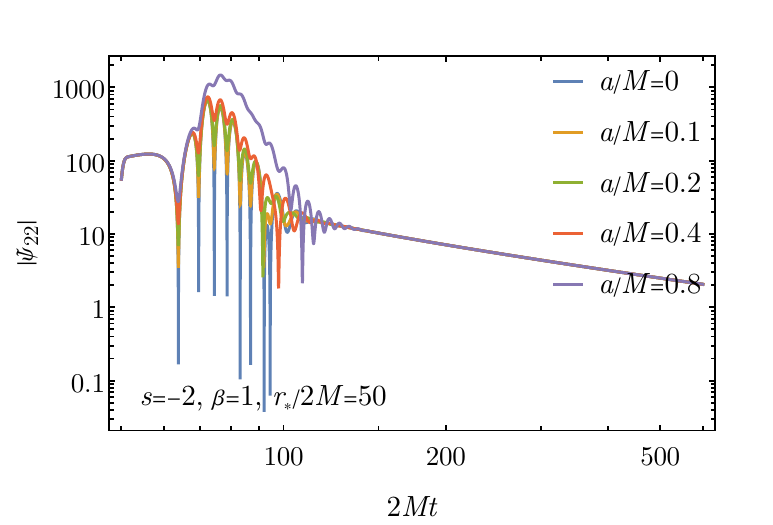} \includegraphics[width=0.48\textwidth]{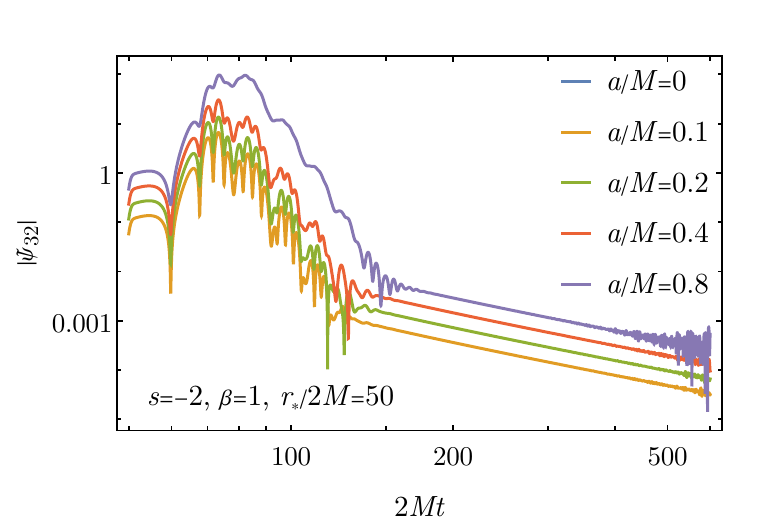}
  \caption{Evolution of $|\tilde{\psi}_{\ell m}(t,r_*)|$ for $(\ell m)=(22), (32)$  and $a/M = 0.1, 0.2, 0.4, 0.8$.
    The tail index for $(\ell m) = (22)$ curves (left panel) is $-1$, and that for $(\ell m) = (32)$ curves are $-2$.
    One can see that changing the rotation parameter $a$ does not affect the tail power law index, but does lead to different early oscillation patterns (quasinormal modes).
  }
  \label{fig:sourced_Psi_loglog_a_comparison}
\end{figure}

Fig.~\ref{fig:sourced_Psi_loglog_a_comparison} illustrates $\abs{\tilde{\psi}}$ evolution for a few different spin parameters $a/M$.
For this set of simulations, the source was taken to be $\tilde{\mathcal{S}} \propto {}_{-2}Y_{2 2}$, so axial symmetry guarantees that the response $\tilde{\psi}_{\ell m}$ is nonzero only for $m=2$.
For $a=0$, spherical symmetry further forbids mode coupling between $(22)$ and other modes, so only $\tilde{\psi}_{22}$ is nonzero.
These symmetry requirements are confirmed by the numerical results.
From the figures, it is also clear that both $\tilde{\psi}_{22}$ and $\tilde{\psi}_{32}$ exhibit power law tails at late times, with the same power law index for different rotation parameters $a/M$.
For $\tilde{\psi}_{22}$, the tails for different $a/M$'s overlap, which corroborates our claim that the tail is primarily sourced far away from the horizon, where the effect of black hole rotation is negligible.
For $\tilde{\psi}_{32}$, the amplitude of the response roughly scales as $(a/M)^1$; this is also expected, since the response (in mode $(32)$) to the source (in mode $(22)$) are connected by couplings that scale like $(a/M)^1$.
Finally, the tail power law for $\psi_{22}$ is $t^{-1}$, and that for $\psi_{32}$ is $t^{-2}$, consistent with the prediction $t^{-\ell-\beta-s}$ with $s=-2$ and $\beta=1$.

\begin{figure}[t]
  \centering
  \includegraphics[width=0.48\textwidth]{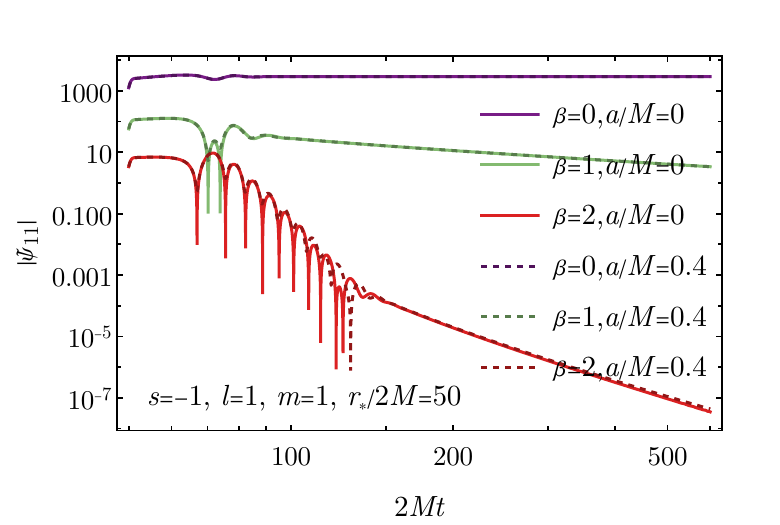}
  \includegraphics[width=0.48\textwidth]{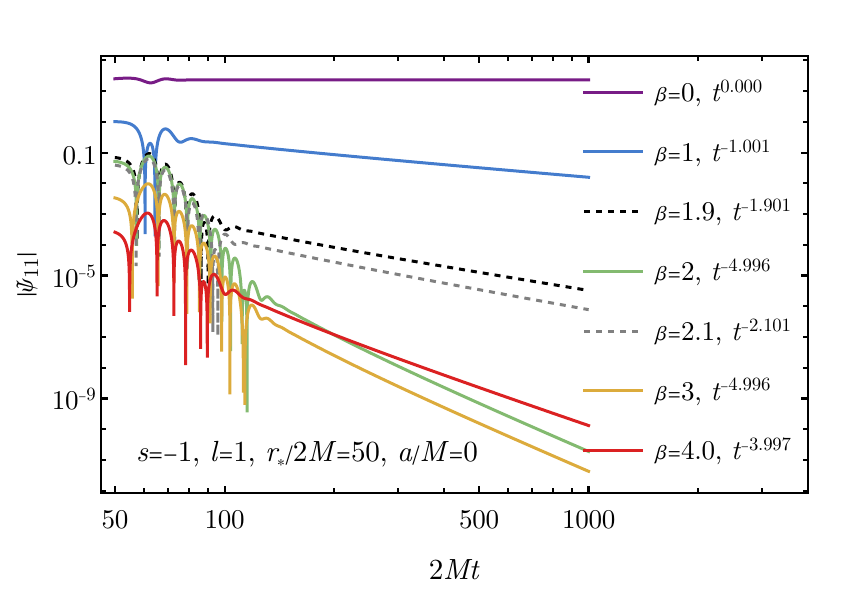}
  
  \caption{Left panel: Evolution of $|\tilde{\psi}_{1 1}(t,r_*)|$ for $\beta = 0, 1, 2$ and $a/M = 0, 0.4$.
    The tail power laws are $t^0$, $t^{-1}$ and $t^{-5}$ for $\beta = 0, 1, 2$, respectively.
    It is clear that rotation do not affect the tail power law indices or amplitudes.
    Right panel: Evolution of $|\tilde{\psi}_{1 1}(t,r_*)|$ for $\beta = 0, 1, 1.9, 2, 2.1, 3, 4$ and $a=0$. The tail power laws are $t^{-\beta}$ for $\beta = 0,1,1.9,2.1,4$, but are $t^{-2\ell-3}$ for $\beta = 2, 3$.
  }
  \label{fig:sourced_Psi_loglog_beta_comparison}
\end{figure}

The left panel of Fig.~\ref{fig:sourced_Psi_loglog_beta_comparison} illustrates $\abs{\tilde{\psi}}$ evolution for a few different source $\beta$'s and rotation parameter $a/M$'s.
For this set of simulations, the source was taken to be $\tilde{\mathcal{S}} \propto {}_{-1}Y_{1 1}$.
Similar to that in Fig.~\ref{fig:sourced_Psi_loglog_a_comparison}, $\tilde{\psi}$ exhibits late time power law tails independent of $a/M$, with the power laws being $t^0$, $t^{-1}$ and $t^{-5}$ for $\beta=0,1,2$, respectively.
For $\beta = 0, 1$, these power law tails again match our analytic prediction $t^{-\ell-\beta-s}$.
While the power laws for $\beta \geq 2$ do not match $t^{-\ell-\beta-s}$, they do match the power law for Price tails~\cite{Krivan:1996da,Barack:1999ma,Barack:1999st}, consistent with prediction \eqref{eq:power_law_prediction}.
Similarly, the right panel of Fig.~\ref{fig:sourced_Psi_loglog_beta_comparison} illustrates the $\abs{\tilde{\psi}}$ evolution for $\beta = 0, 1, 1.9, 2, 2.1, 3, 4$ and $a/M=0$. It was found that the tail is consistent with the predicted $t^{-\ell-\beta-s}$ for $\beta = 0, 1, 4$ and non-integer $\beta$'s, but exhibits the Price power law $t^{-2\ell-3}$ for $\beta = 2, 3$. These results are in exact agreement with analytical prediction \eqref{eq:power_law_prediction}, which states that the Price tail dominates for $2 \leq \beta \leq \ell - s + 1$. Notably, sharp jumps in the tail power law and amplitudes appear when $\beta$ is varied across $2$ and $3$. For example, although the tail power law for $\beta = 1.9$ and $2.1$ is respectively $t^{-1.9}$ and $t^{-2.1}$, it sharply drops to $t^{-5}$ for $\beta = 2$.

Table.~\ref{table:power_laws} provides the extrapolated power laws for $a=0$ sourced simulations.
The $s = 0$ results are identical with the ones found in Ref.~\cite{Ling:2025wfv}.
For $\beta = 0, 1$ and $s < 0$, the tail power law is $t^{-\ell-\beta-s}$, consistent with \eqref{eq:power_law_prediction}.
For $2 \leq \beta \leq \ell - s + 1$ and $s < 0$, the tail generally decays according to the Price law $t^{-2\ell-3}$, again consistent with our analytical prediction \eqref{eq:power_law_prediction}.

\begin{table}[h!]
  \centering
  \begin{tabular}{ |c|c|c| }
    \hline
    \multicolumn{3}{|c|}{\Revise{$s=0$}} \\
    \hline
    & $\beta = 0$ & $\beta = 1$
    \\ \hline
    $\ell=1$&$-1.00$&$-1.99$\\$\ell=2$&$-2.00$&$-2.99$
    \\ \hline
  \end{tabular}
  \begin{tabular}{ |c|c|c|c| }
    \hline
    \multicolumn{4}{|c|}{\Revise{$s=-1$}} \\
    \hline
    & $\beta = 0$ & $\beta = 1$ & $\beta = 2$
    \\ \hline
    $\ell=1$&$0.00$&$-1.00$&$-5.00$\\$\ell=2$&$-1.00$&$-2.00$&$-6.99$
    \\ \hline
  \end{tabular}
  \begin{tabular}{ |c|c|c|c| }
    \hline
    \multicolumn{4}{|c|}{\Revise{$s=-2$}} \\
    \hline
    & $\beta = 0$ & $\beta = 1$ & $\beta = 2$
    \\ \hline
    $\ell=2$&$-0.00$&$-1.00$&$-7.00$
    \\ \hline
  \end{tabular}
  \caption{Extrapolated power laws for different $(s, \ell, \beta)$ parameters.
    \Revise{Each table corresponds to a fixed $s$.}
    The extrapolation was performed by fitting the local power law $t \partial_t \ln |\tilde{\psi}|$ to the template $p(t) = a + b / (t - c)$ over $t \in [200,800]$, and the reported power laws are the fitted value for $a$.
  } 
  \label{table:power_laws}
\end{table}

\subsection{Scalar perturbation with cubic interaction}
\label{sec:scalar_result}
The previous section tested our analytic understanding at the level of perturbation theory. To further assess whether this picture of nonlinear tails remains valid in fully nonlinear systems, the ideal case would be to use data from numerical relativity. As a prototype, however, we consider the evolution of a scalar perturbation with a cubic interaction, similar to the setup in Ref.~\cite{Ling:2025wfv}, with Kerr spacetime background instead of Schwarzschild.
Precisely, we take real scalar field $\psi$ that satisfies
\begin{align}
  \label{eq:cubic_scalar_eqn_covariant}
  0 = -\nabla^a \nabla_a \psi + \lambda \psi^2 \;.
\end{align}
Expanding the above equation in Kerr metric reveals that $\psi$ satisfies the Teukolsky equation \eqref{eq:teukolsky} with spin $s=0$.
We use the change of variable $\tilde{\psi} \equiv r \psi$ discussed in Sec.~\ref{sec:teukolsky_equation}, and set the initial condition as  an ingoing Gaussian wavepacket:
\begin{align}
  \label{eq:cubic_scalar_Psi_IC}
  &\tilde{\psi}_{11}(t=0,r_*) = \frac{1}{\sigma \sqrt{2\pi}} e^{-\frac{(r_*-r_s)^2}{2 \sigma^2 }},\quad
                       \partial_t\tilde{\psi}_{11}(t=0,r_*) = \partial_{r_*} \tilde{\psi}_{11}(t=0,r_*) \nonumber \\
  &  \sigma/2M = 0.5,\quad r_s/2M = 50 \nonumber \\
  & \tilde{\psi}_{\ell m}(t=0,r_*) = \partial_t \tilde{\psi}_{\ell m}(t=0,r_*) = 0 \qq{for} (\ell m) \neq (11) \;.
\end{align}
The $a=0$ case of this setup is exactly Eq.~(3.3) of Ref.~\cite{Ling:2025wfv}.

In Ref.~\cite{Ling:2025wfv}, a perturbative method was used to study Eq.~\eqref{eq:cubic_scalar_eqn_covariant} on Schwarzschild background: the full solution $\psi = \sum_k \lambda^k \psi^{(k)}$ is arranged order-by-order in $\lambda$, with lower order solutions acting as sources for higher order solutions.
A similar treatment can be applied to Eq.~\eqref{eq:cubic_scalar_eqn_covariant} on Kerr backgrounds.
In contrast to the Schwarzschild case, additional couplings between harmonic modes arise at linear order due to the breakdown of spherical symmetry.
In the Teukolsky equation \eqref{eq:psi_lm_pde}, there are generally couplings between modes $(\ell m)$ and $(\ell' m')$ for $m=m'$, with the coupling coefficients scaling like power laws of $a/M$ for $\abs{a/M} \ll 1$.
Furthermore, for $s=0$, parity symmetry forbids couplings between modes with odd $\ell-\ell'$.
Given our initial condition in $(11)$, the modes excited at $\lambda^0$ level are $(11)$, $(31)$ and $(51)$.
The modes excited at $\lambda^1$ level are $(22)$ and $(42)$, and at $(33)$ and $(53)$ for $\lambda^2$.

\begin{figure}[t]
  \centering
  \includegraphics[width=0.48\textwidth]{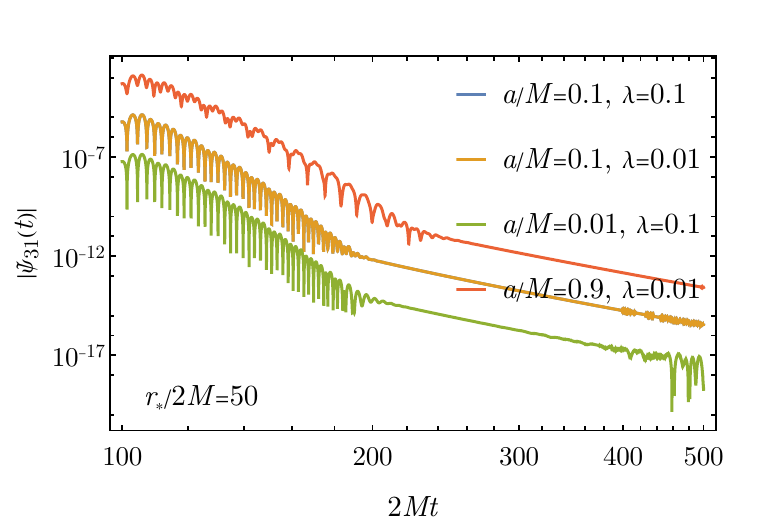} \includegraphics[width=0.48\textwidth]{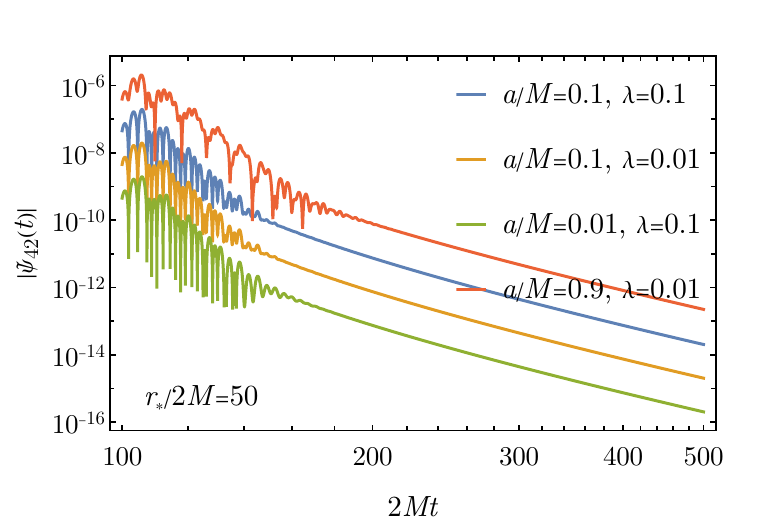}
  \caption{Comparison of tails dynamically formed due to a cubic coupling.
    The power laws for the curves in the left (right) panel is $t^{-4}$ and $t^{-5}$, consistent with the $t^{-\ell-1}$ analytical prediction.
    On the left panel, the blue and yellow curves overlap, and the amplitudes scale like $(a/M)^2$.
    On the right panel, the amplitudes scale like $\lambda^1 (a/M)^2$.
  }
  \label{fig:coupled_Psi_tail_comparison}
\end{figure}

Fig.~\ref{fig:coupled_Psi_tail_comparison} shows the evolution of $\tilde{\psi}_{42}$ for some choices of $a/M$ and $\lambda$.
It is clear that the power law tail for fixed $\ell$ does not depend on the $a/M$ or $\lambda$, and the amplitude does scale predictably with $a/M$ and $\lambda$.
Across all $(\ell m)$ modes, the tail power law is always $t^{-\ell-1}$, consistent with the $t^{-\ell-\beta}$ prediction (for $\beta=1$) and identical with the Schwarzschild case.
For $\tilde{\psi}_{42}$, the amplitude scales as $\lambda^1$ and $(a/M)^2$, consistent with the perturbative analysis.
In summary, for this cubic scalar field model, the nonlinear tail power law for Kerr black hole ringdown is always the same as that for Schwarzschild black hole.

\section{Discussion \& Conclusion}
\label{sec:discussion}

In this work, we generalize the analysis of nonlinear tail power laws in Ref.~\cite{Ling:2025wfv} from Schwarzschild to Kerr black holes.
Our framework is based on the Teukolsky equation with arbitrary spin weight $s$, and therefore applies uniformly to scalar, electromagnetic, and gravitational perturbations. 
Analytically, we derive the far field radial Green's function and obtain a sharp prediction for the nonlinear tail power law, Eq.~\eqref{eq:power_law_prediction}, generated by outgoing sources that decay as $r^{-\beta}$.
Numerically, we validate this picture in two complementary settings: first, by directly evolving the response to an explicit outgoing source (Fig.~\ref{fig:sourced_Psi_loglog_a_comparison}), and second, by demonstrating the dynamical emergence of nonlinear tails in simulations of a cubically self-interacting scalar field.
Our central result is that the late-time nonlinear tails of Kerr black holes obey the same power laws as in the Schwarzschild case.
In particular, for $\beta \leq 1$ and $\beta \geq \ell - s + 2$, both analytic results and numerical simulations consistently yield the universal scaling $t^{-\ell-\beta-s}$.
Although our analysis was performed for toy models, we expect our analysis to capture the qualitative features of GR nonlinear tails.

Both our numerical and analytical results indicate that Kerr black hole nonlinear tails have the same power law indices as that for Schwarzschild black holes; this holds even for black holes with significant angular momentum, for different spin-weights, and for each harmonic mode.
The phenomenon has a clear physical interpretation: nonlinear tails are sourced far away from the black hole, and are thus only sensitive to the spacetime background in that regime.
Physically speaking, since both Kerr and Schwarzschild spacetime are asymptotically Minkowski at large $r$ ($\abs{a}/r \ll 1$ and $M/r \ll 1$), the sourced nonlinear tails in the two cases should share common behaviors.
Additionally, if the source is itself dynamically propagating on Kerr background, its propagation is also only sensitive to the far field regime; this justifies a source of the form \eqref{eq:outgoing_source}.

When $a \neq 0$, couplings between spin-weighted spherical harmonic modes $(\ell m)$ of the same magnetic quantum number $m$ arise~\cite{Hod:1999rx,Hod:1999ry,Hod:2000fh,Burko:2013bra}, leading to nonlinear tails in modes distinct from the source mode.
As illustrated in Fig.~\ref{fig:sourced_Psi_loglog_a_comparison}, these couplings and nonlinear tails scale predictably with $a/M$.
Interestingly, the tail power law for all harmonic modes seem to be consistent with the ones obtained in the Schwarzschild case, even when the modes are sourced due to Kerr-only couplings.
Further work would be required to understand this fact, and whether the $r$ dependence of the Kerr-only couplings play a role in determining the tails.
Finally, since higher $\ell$ modes have faster power law decays than lower $\ell$ modes, the lower $\ell$ modes are still expected to dominate the late time nonlinear tails of a black hole.

Although the Teukolsky equation provides a technically complete description of Kerr spacetime perturbations, understanding the observational consequences of its solutions is not a trivial task.
For example, the Teukolsky equation governs evolution of Weyl scalar $\psi_4$ (which has the same power law dependence on $t$ as $\tilde{\psi} \equiv (\Delta^s r) \psi$), and reconstructing metric perturbations from the Weyl scalars is known to be a cumbersome task~\cite{Ori:2002uv}.
However, physical predictions do not depend on the choice of formalism, so the conclusion that Kerr black hole nonlinear tails have the same power laws as Schwarzschild black holes is still solid.
In fact, recent numerical studies in Ref.~\cite{Bishoyi:2026cpg} confirmed that Kerr black holes have decay power law $t^{-2\ell-2}$, consistent with the analytical predictions based on the Regge-Wheeler equation~\cite{Kehagias:2025xzm}.
Another nuance is that observable tails were shown to exhibit power laws evaluated at null infinity instead of at future infinity~\cite{Zenginoglu:2008wc,Gundlach:1993tp}.
In our context, this implies the observed power law should be $u^{-\ell-1}$ instead of $t^{-2\ell-2}$.

We conclude by highlighting two closely related directions for future work.
First, it would be useful to connect nonlinear tails to concrete astrophysical channels that can generate strong outgoing disturbances far from the black hole.
In particular, hyperbolic encounters in compact-object systems, as well as energetic episodes such as gamma-ray outbursts from accreting black holes, may act as physically relevant sources of nonlinear tails.
A key question then is whether the resulting late-time signal carries observable information about the source dynamics, and how such information can be disentangled from the underlying linear response and the ordinary Price tail.
Second, the excitation of nonlinear tails is likely tied to the broader problem of nonlinear mode coupling in black hole ringdown.
In this perspective, the detailed pattern of mode excitation in nonlinear ringdown~\cite{Ma:2024qcv} may offer a useful window into how nonlinear tails are seeded and amplified; conversely, tail measurements may provide indirect information about the nonlinear transfer of energy across modes.
Clarifying these connections would help place nonlinear tails within a more complete physical picture of black hole dynamics and their potential observational signatures.

\begin{acknowledgments}
 We thank Anil Zenginoglu for interesting discussions on the distinction between null infinity and future infinity power laws. \Revise{This project is supported by grant No. 9610650 from APRC-CityU New Research Initiatives/Infrastructure Support from Central and ECS Grants No. 21313926 from the Research Grants Council, University Grants Committee.}
  The data that support the findings of this study were generated by numerical simulations. The source code and parameters used to generate the simulations is publicly available~\cite{BlackholePerturbations}.
  
\end{acknowledgments}

\appendix

\section{Numerical implementation}
\label{sec:numerical_implementation}
Here we describe our numerical implementation for solving the generic Teukolsky equation as decoupled 1+1D PDEs.

\subsection{Numerical setup}
For all our simulations, we set a cutoff $\ell_{\mathrm{max}} = 5$ and include all harmonics with $\ell \leq \ell_{\mathrm{max}}$.
Each simulation thus consists of $(\ell_{\mathrm{max}}+1)^2 = 36$ coupled 1+1D PDEs (including harmonics that never get sourced), forming a closed system of equations.
The spatial resolution was $\Delta{r}_* = 0.03$ over domain $r_* \in [-500,1000]$, and 2nd order spatial derivatives were approximated using a 4th order finite difference scheme.
Time steps were fixed to $\Delta{t} = 0.01$, with time evolution performed using an 5th order Runge-Kutta method.
Double precision floating point numbers were used for these simulations.
With these settings, we evolved the equations from $t=0$ to $t=1000$.
Our domain and time limits are chosen such that reflections from spatial boundaries never reach the observation point $r_*=50$, so reflections have no effect on extracted tails.
All numerical evolution were written in CUDA and ran on NVIDIA GPUs.
To reduce kernel launch overheads and to parallelize the workload for different spherical harmonic modes, CUDA graph was used to arrange the computation.
Depending on the setting, each simulation took from 10 to 20 minutes on a RTX 4070 graphics card, a significant speed up compared to CPU implementations.
For verification, we performed convergence tests with shorter time step ($\Delta{t} = 0.005$), with higher spatial resolution ($\Delta{r}_* = 0.015$), and with different Kreiss-Oliger dissipation parameters, and we found no qualitative difference in the results.
In addition, we performed $a=0$ runs using quadruple precision floating point numbers on CPU, and found the results to exhibit lower numerical noise, but are otherwise identical to those obtained from double precision floating point numbers.
Our code is released at \url{https://github.com/hypermania/BlackholePerturbations}.

\subsection{Multiplication in spherical harmonic space}

Suppose we have two scalar fields $f$ and $g$, and we want to express the quadratic term $f g$ in terms of the spherical harmonic components of $f$, we have:
\begin{align}
  \label{eq:spherical_harmonic_product}
  f &= \sum_{\ell m} f_{\ell m} Y_\ell^m,\quad
      g = \sum_{\ell m} g_{\ell m} Y_\ell^m \nonumber \\
  (f g)_{\ell^{\prime\prime} m^{\prime\prime}} &= \sum_{\ell m} \sum_{\ell' = \abs{\ell^{\prime\prime} - \ell}}^{\ell^{\prime\prime}+\ell} f_{\ell m} g_{\ell', m^{\prime\prime} - m} E\indices{_0^0_{\ell'}^\ell_{m^{\prime\prime}-m}^m_{\ell^{\prime\prime}}} \;,
\end{align}
where $E$ is the Edmond symbol, defined in equation (38) of Ref.~\cite{Brizuela:2006ne}.
It is the coefficient of the product of two scalar spherical harmonics $Y_\ell^m$:
\begin{align}
  Y_{\ell'}^{m'} Y_\ell^m &= \sum_{\ell^{\prime\prime} = \abs{\ell'-\ell}}^{\ell'+\ell} E\indices{_0^0_{\ell'}^\ell_{m'}^m_{\ell^{\prime\prime}}}
                      Y_{\ell^{\prime\prime}}^{m+m'} \nonumber \\
  E\indices{_0^0_{\ell'}^\ell_{m'}^m_{\ell^{\prime\prime}}} &= \sqrt{\frac{(2\ell+1)(2\ell'+1)}{4\pi (2\ell^{\prime\prime}+1)}} C\indices{_{\ell'}^{m'}_\ell^m_{\ell^{\prime\prime}}^{m'+m}} C\indices{_{\ell'}^0_\ell^0_{\ell^{\prime\prime}}^0} \;,
\end{align}
where $C$ are the Clebsch-Gordan coefficients.
Eq.~\eqref{eq:spherical_harmonic_product} provides a numerical recipe for evaluating products of scalar fields in terms of mode coefficients, and it can be extended to any number of products by repeated application.

For Teukolsky equation, the cubic term is of the form $c \psi^2$, where $c$ is a $\theta$ dependent coefficient that can be expanded in terms of spherical harmonics explicitly:
\begin{align}
  (c \psi^2)_{\ell m}
  &= \sum_{\ell' m'} \sum_{\ell^{\prime\prime}=\abs{\ell-\ell'}}^{\ell+\ell'} c_{\ell'm'} (\psi^2)_{\ell^{\prime\prime}, m-m'} \;.
\end{align}
Numerical implementation of such a product can be done in two stages using Eq.~\eqref{eq:spherical_harmonic_product}: simply compute $(\psi^2)_{\ell m}$, and multiply the product with $c_{\ell m}$.

\subsection{Kreiss-Oliger dissipation}
During numerical evolution of the Teukolsky equation, highly oscillatory modes (those at the Nyquist limit) will generally grow exponentially due to numerical instability.
An artificial dissipation term, known as the Kreiss-Oliger dissipation, is added to the numerical evolution to suppress these modes.
Schematically, the numerical evolution is given by
\begin{align}
  \partial_t \mqty[\tilde{\psi}\\ \partial_t \tilde{\psi}] = M \mqty[\tilde{\psi}\\ \partial_t \tilde{\psi}] + \mqty[- Q \tilde{\psi} \\ \tilde{\mathcal{S}}] \;,
\end{align}
where $M$ is defined by the Teukolsky operator $\tilde{\mathcal{T}}$, and $Q$ is the Kreiss-Oliger (KO) operator, a 4-th order spatial derivative:
\begin{align}
  (Q\psi)_i = \frac{\epsilon}{16} ( \psi_{i-2} - 4 \psi_{i-1} + 6 \psi_{i} - 4 \psi_{i+1} + \psi_{i+2}) \;.
\end{align}
This operator scales as $(\Delta{r}_*)^4$ in amplitude and is proportional to the 4-th spatial derivative where $\tilde{\psi}$ is spatially smooth.
We performed numerical checks and found that larger spin-weights typically require higher $\epsilon$ for convergence.
In our numerical simulations, we chose $\epsilon = 0.7$, which is adequate for all spin-weights $\abs{s} \leq 2$.

\section{Far field Green's function derivation}
\label{sec:greens_function_derivation}

Here we derive an approximation to the Green's function in the far field limit.
In the far field ($M / r \ll 1$, $\abs{a} / r \ll 1$) limit, the homogeneous equation simplifies to
\begin{align}
  0 &= \pdv[2]{{}_s\tilde{\psi}_{\ell m}}{t} + \frac{2 s}{r} \pdv{{}_s\tilde{\psi}_{\ell m}}{t}
      - \left( \pdv[2]{r} - \frac{2 s}{r} \pdv{r} \right) {}_s\tilde{\psi}_{\ell m}
      + \frac{1}{r^2} (\ell-s)(\ell+s+1) {}_s\tilde{\psi}_{\ell m}  \;.
\end{align}
In frequency space ${}_s\tilde{\psi}_{\ell m}(\omega, r)$, the above becomes
\begin{align}
  \label{eq:psi_tilde_eqn_frequency}
  0 &= \pdv[2]{{}_s\tilde{\psi}_{\ell m}}{r} - \frac{2 s}{r} \pdv{{}_s\tilde{\psi}_{\ell m}}{r} 
  - \frac{1}{r^2} (\ell-s)(\ell+s+1) {}_s\tilde{\psi}_{\ell m}
  - (- \omega^2 - i \omega \frac{2s}{r}) {}_s\tilde{\psi}_{\ell m} \;.
\end{align}
We use the usual frequency space approach to construct the Green's function~\cite{Leaver:1986gd,Ling:2025wfv} (for convenience we temporarily suppress the subscripts $s \ell m$) 
\begin{align}
  \label{eq:G_psi_contour_integral}
  G(r, t; r', t')  &= \frac{1}{2 \pi i} \int_{ic - \infty}^{ic + \infty} e^{- i \omega (t-t')} g(r,r',\omega) (- i) \dd{\omega} \nonumber \\
g(r,r',\omega) &\equiv \frac{\tilde{\psi}_{\infty_-}(\omega, r_<) \tilde{\psi}_{\infty_+}(\omega, r_>) }{W[\tilde{\psi}_{\infty_+}, \tilde{\psi}_{\infty_-}](r')} \nonumber \\
  \tilde{\psi}_{\infty_+}(\omega, r) &= r^s W_{-s,\ell+1/2}(-2 i \omega r) \nonumber \\
  \tilde{\psi}_{\infty-}(\omega, r) &= r^s W_{s, \ell+1/2}(2 i \omega r) \nonumber \\
  W[\tilde{\psi}_{\infty_+}, \tilde{\psi}_{\infty_-}](r) &= (-1)^{s+1} 2 i \omega r^{2s} \;,
\end{align}
where $r_> \equiv \mathrm{max}(r,r')$, $r_< \equiv \mathrm{min}(r,r')$, $W_{k,m}(z)$ is the Whittaker function, $W[\cdot, \cdot]$ is the Wronskian, and $c > 0$ is chosen such that the integration contour is above all poles in the complex $\omega$ plane.
One can check that $\tilde{\psi}_{\infty_\pm}$ are indeed exact solutions of \eqref{eq:psi_tilde_eqn_frequency}; in fact, they are the outgoing and ingoing solutions at $r \to \infty$ and asymptote to $\tilde{\psi}_{\infty_\pm}(\omega, r) \sim e^{\pm i \omega r} r^{s \mp s}$.
For $t-t' > \abs{r_*-r_*'}$, the integration contour is closed in the lower half plane and can be deformed to a counterclockwise loop around $\omega = 0$; otherwise, the contour is closed in the upper half plane, and the integral is zero.
This distinction gives rise to a theta function $\Theta(t-t'-\abs{r_*-r_*'})$, which enforces causality.
For specific $(s, \ell, m)$, the contour integral for $G$ can be now be readily computed using the residue theorem.
Unlike the case for Price tail calculations, branch cut contributions vanish in the far field approximation and are absent from above.

Let $\bar{G}_{s \ell m}$ be the Green's function without the causal theta function, $G_{s \ell m} = \bar{G}_{s \ell m} \Theta(t-t'-|r_*-r_*'|)$.
We list some results for the Green's function approximations derived in the far field limit:
\begin{align}
  \bar{G}_{s (\ell = 0) m}(r,t;r',t') &= \frac12 \qquad s=0 \nonumber \\
  \bar{G}_{s (\ell = 1) m}(r,t;r',t') &=
                                  \begin{cases}
                                    \frac{(r + r' + \Delta{t})^2}{8 r^2} & s = -1 \\
                                    \frac{r^2 + (r')^2 -\Delta{t}^2}{4 r r'} & s = 0 \\
                                    \frac{(r + r' - \Delta{t})^2}{8 (r')^2} & s = 1 
                                  \end{cases} \nonumber \\
  \bar{G}_{s (\ell = 2) m}(r,t;r',t') &=
                                  \begin{cases}
                                    \frac{(r + r' + \Delta{t})^4}{32 r^4} & s = -2 \\
                                    \frac{(r + r' + \Delta{t})^2(r^2 - r r' + (r')^2 - (\Delta{t})^2)}{8 r^3 r'}  & s = -1 \\
                                    \frac{3 r^4 + 2 r^2 ((r')^2 - 3 \Delta{t}^2) + 3( (r')^2 - \Delta{t}^2)^2}{16 r^2 (r')^2} & s = 0 \\
                                    \frac{(r + r' - \Delta{t})^2(r^2 - r r' + (r')^2 - (\Delta{t})^2)}{8 r (r')^3} & s = 1 \\
                                    \frac{(r + r' - \Delta{t})^4}{32 (r')^4} & s = 2
                                  \end{cases}
\end{align}
where $\Delta{t} \equiv t-t'$.
As proved in Ref.~\cite{Ling:2025wfv}, causality considerations imply that this approximation is only valid for $\abs{r_* - r_*'} < t-t' < r_* + r_*'$.
For $s = 0$, these Green's functions are identical to the Green's function approximations derived for the Regge-Wheeler equation in Ref.~\cite{Barack:1999ma,Ling:2025wfv}; our work generalizes those results to $s \neq 0$.
We have also computed Green's functions via direct numerical integration, and found that the above results are accurate within their regime of validity.
Note that these Green's functions satisfy the spin-weight inversion identity.

\bibliography{main,data_ref}

\end{document}